\documentclass[12pt]{article} 
\usepackage{amsmath}
\usepackage{amssymb}
\usepackage{latexsym}
\usepackage{amsthm}
\usepackage{amsbsy}
\usepackage{amstext}
\usepackage[dvips]{graphicx}

\def\dis{\displaystyle}
\def\r{\boldsymbol{r}}
\def\p{\partial}
\begin{document} 
%

\vspace{10mm}

\begin{center}
{\Large \bf 
In-situ observation of a soap film catenoid -
a simple educational physics experiment}\\
\vspace{20mm}
 Masato Ito
 \footnote{mito@auecc.aichi-edu.ac.jp} and Taku Sato
\end{center}

\begin{center}
{
{}Department of Physics, Aichi University of Education, Kariya, 
448-8542, JAPAN
}
\end{center}

\vspace{25mm}

\begin{abstract}
 The solution to the Euler-Lagrange equation is an extremal functional.
 To understand that the functional is stationary at local extrema
 (maxima or minima), we propose a physics experiment that involves using
 soap film to form a catenoid.

 A catenoid is a surface that is formed between two coaxial circular
 rings and is classified mathematically as a minimal surface.
 Using soap film, we create catenoids between two rings and
 characterize the catenoid in-situ while varying distance between rings.
 The shape of the soap film is very interesting
 and can be explained using dynamic mechanics.
 By observing catenoid, physics students can observe local extrema
 phenomena. 
 We stress that in-situ observation of soap film catenoids is an
 appropriate physics experiment that combines theory and experimentation.
\end{abstract}
\newpage
\section{Introduction}

The solution to the Euler-Lagrange equation is an extremal functional,
which is stationary at local extrema (maxima or minima).
For example, in classical mechanics
\cite{book5}, the functional (Lagrangian) for
any given system leads to Newton's law of motion.
However, physics students may have difficulty recognizing
the behaviour of extremal functionals in physics experiments.
Although we can observe some physical phenomena that are solutions to
the Euler-Lagrange equation, the observation of local extrema is not easy.
We therefore propose an accessible experiment that allows students to
observe local extrema of a minimal surface, which is a solution to the
Euler-Lagrange equation when choosing surface area as the functionals
\cite{book1,book2,book3,book4,book6,book7}.

We stress that experimenting with the minimal surface
formed by soap film is very interesting.
In particular, the shape of a soap film inside a wire frame minimizes
the surface area.
The surface is classified mathematically as a minimal surface and,
from the view point of dynamics and according to the principle of least
action, its shape results from minimizing the surface
energy of the soap film.

In the proposed experiment, we observe the dynamic behaviour of a
catenoid, which is the shape a soap film takes between two coaxial
rings.
The catenoid can be explained using dynamic mechanics and differential
geometry.
We propose that in-situ observation and mathematical analysis of a
catenoid soap film is suitable for an educational curriculum in
experimental physics.

This article is organized as follows.
In Sec.II, we describe the mathematical properties of the minimal
surface and analyse the the shape of the catenoid.
In Sec.III, we explain how to observe the catenoid and present
some results of in-situ observation of a catenoid.
In Sec.IV, we interpret the behaviour of the catenoid using
dynamical mechanics.
Finally, we give a brief summary and conclude in Sec.V.

\section{Mathematics of Catenoid}

In this section, we explain the mathematics of a minimal surface.
For a given boundary, the minimal surface is that which has an extremal
area \cite{book1,book2,book4}.
If a minimal surface is parametrised in a three-dimensional
orthogonal coordinate system $(x,y,f(x,y))$, the 
Euler-Lagrange equation leads to
\begin{align}
\left(1+f^{2}_{y}\right)f_{xx}+\left(1+f^{2}_{x}\right)f_{yy}
-2f_{x}f_{y}f_{xy}=0\,,
\end{align}
where $f_{i}$ denotes the partial derivative with respect to index
$i$.
Because the partial differential Eq. (1) is rather complicated, it is
difficult to solve.
However it can be easily solved by imposing some assumptions
(e.g., symmetry or boundary condition).
Some well-known minimal surfaces include the catenoid, Helicoid, Scherk's
surface, Enneper's surface, etc \cite{book1,book2,paper1,paper2,paper3}.
In this study, we focus on the catenoid, which
is the minimal surface between two coaxial circular rings. 

If the minimal surface described by Eq. (1) is restricted to a surface of
revolution with axial symmetry, we obtain the equation of a catenoid
\cite{book1,book2,paper1,paper2,paper3}.
As shown in Fig. \ref{fig-1}, a catenoid is a
surface of revolution generated from a catenary curve with radius
\begin{align}
r(z)=a\cosh\frac{z}{a}\,,
\end{align}
where $a$ is the minimum value of $r$ at $z=0$.
In the appendix, we derive Eq. (2) using the Euler-Lagrange equation.

\begin{figure}
\begin{center}
\includegraphics[width=12cm]{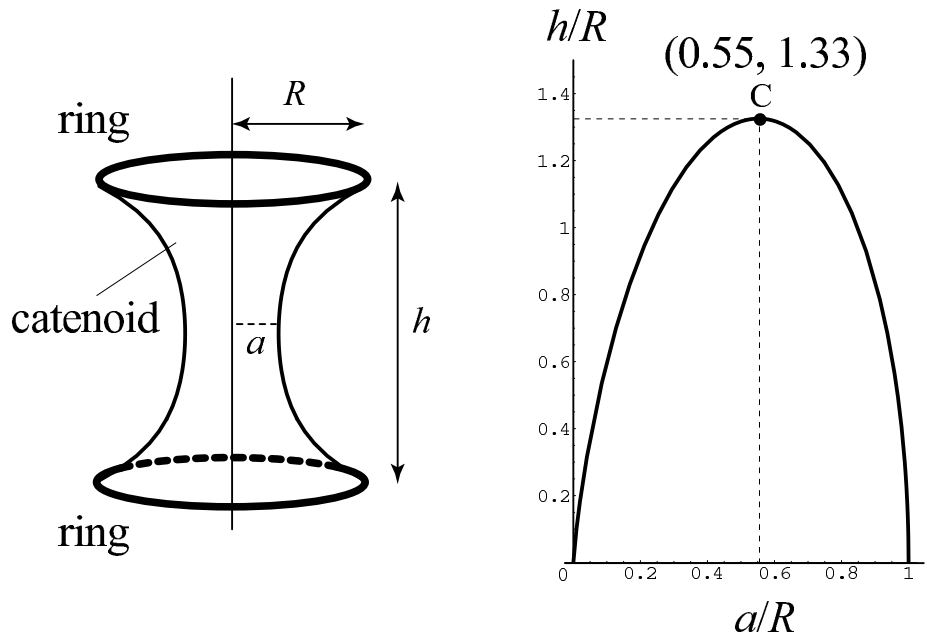}
\caption{\label{fig-2}
Drawing of a catenoid showing the three parameters
(the distance $h$ between the two circular rings, the radius $R$ and
the neck radius $a$)
on which the catenoid shape depends (left panel).
The graph of Eq. (3) is shown (right panel).
The point C corresponds to the maximum inter-ring separation.
}
\end{center}
\end{figure}

\begin{figure}
\begin{center}
\includegraphics[width=11cm]{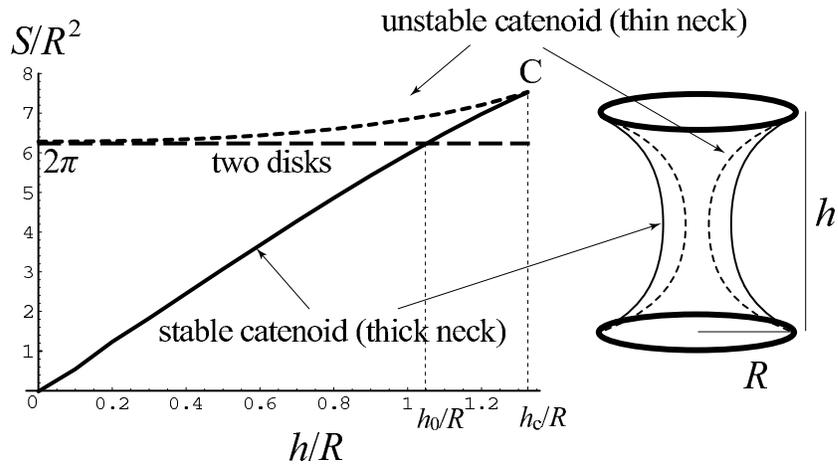}
\caption{\label{fig-3}
The surface area $S$ plotted as a function of the distance $h$ between
the two rings.
The distance corresponding to point C is $h_{c}\simeq 1.33 R$.
The intersection between the stable catenoid curve and the horizontal
dashed line corresponding to two disks occurs at $h_{0}\simeq 1.05 R$.
}
\end{center}
\end{figure}

In Fig. \ref{fig-2}, the catenoid is described by three parameters
$h,R$ and $a$, where
$h$ is the distance between two coaxial circular rings, $R$ is
the radius of the equal-sized rings and
$a$ is the neck radius. 
According to Eq. (2), three parameters $h,R$ and $a$ are related by
\begin{align}
h=2a\cosh^{-1}\frac{R}{a}\,.
\end{align}
As indicated by the right-handed graph of Fig. \ref{fig-2},
the shape of a catenoid depends on the ratio $h/R$.
For $h/R>1.33$, it is not possible to form a catenoid, so
the critical distance $h_{c}$ between rings is
$h_{c}\simeq 1.33 R$.
For any given $h<h_{c}$, two catenoids are possible.
Figure {\ref{fig-2}} shows that the neck radius of one catenoid (thick neck) is
larger than the neck radius of other catenoid (thin neck).
However, only one catenoid can physically exist, because the
two catenoids correspond to extremal surfaces with a local maximum or
a local minimum, we must compare the surface areas of two catenoids
to determine which can exist.
Using differential geometry, we obtain the following expression for the 
surface area $S$ of the catenoid as a function of $a$:
\begin{align}
S=2\pi a^{2}\left(
\cosh^{-1}\frac{R}{a}+\frac{1}{2}\sinh\left(2\cosh^{-1}\frac{R}{a}\right)
\right)\,.
\end{align}
Using Eqs. (3) and (4), the graph of $S$ and $h$ are depicted in 
Fig. \ref{fig-3} by Mathematica 5.1.
As shown in Fig. \ref{fig-3}, the surface area of the thick-neck
catenoid is smaller everywhere than that of the thin-neck catenoid, so
the surface area of the thick-neck catenoid is the absolute
minimum for all values of $h$.
Thus, the thick-neck catenoid is dynamically stable and the thin-neck
catenoid is unstable.
In addition, we must consider the area of the disk inside the ring.
Thus, we added a horizontal dashed line in Fig. \ref{fig-3},
which corresponds to the surface area of the two disks.
When the surface area of a stable catenoid equals the surface area of
the two disks, the rings are separated by $h_{0}\simeq 1.05 R$.

We are interested how the shape of the catenoid behaves as a function of
the ring-separation $h$.
Since the absolute minimum surface area is selected, the surface area of
catenoid must always have the absolute minimum value.
It is likely that a majority of physics students agree with this idea.
From the viewpoint of classical mechanics, it seems that some physical
phenomena in the natural world are described by functionals with
absolute minimum.

From the viewpoint of the absolute minimum,  
we can predict the catenoid shape while gradually increasing $h$.
As shown in Fig. \ref{fig-3}, for $0<h< h_{0}$, a stable
thick-neck catenoid exists.
For $h_{0}<h< h_{\rm c}$, two disks exist.
At $h=h_{0}$, the stable catenoid jumps to the two disks.
Thus, we expect that the stable catenoid separates in two and transfers
to the two rings.

To confirm if the idea is correct, we performed the soap-film
experiment.
By increasing the distance between the two support rings, we can
determine if the soap film changes from a stable catenoid to two rings
at $h=h_{0}$.
In the next section, we show the results of this experiment.

\section{in-situ observation of a catenoid}

In the middle of the 19th century, physicist J. Plateau proposed
that minimal surfaces could be visualized using soap film.
To see the minimal surface for a given arbitrary boundary,
it sufficed to soak a wire frame with its boundary in soapy
water \cite{book2}.
The soap film occupies the minimal surface to help
minimize the surface tension.
This demonstration, which combined mathematics and physics, was a great
achievement by J. Plateau.

\begin{figure}
\begin{center}
\includegraphics[width=8cm]{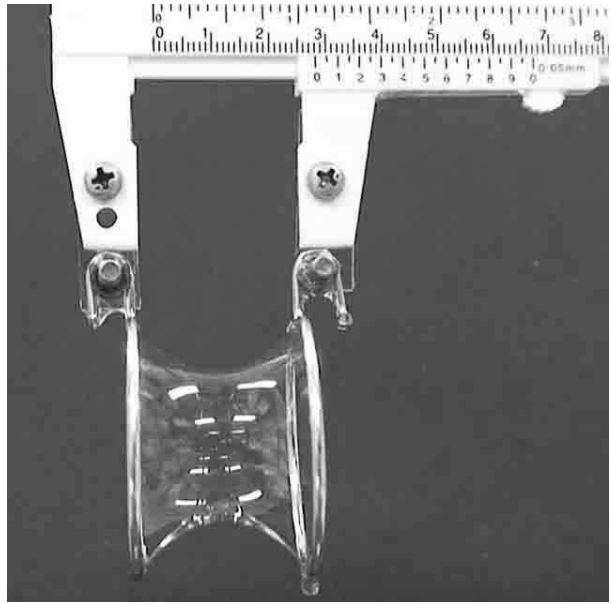}
\caption{\label{fig-4}
Photo of a slide caliper with two circular rings making a
soap-film catenoid.}
\end{center}
\end{figure}

To observe the shape of the soap film,
we used a familiar slide caliper with circular rings attached to the
tips of the caliper (Fig. \ref{fig-4})
The device enabled us to measure simultaneously both the distance $h$
between the two rings and the neck radius $a$.

\begin{figure}
\begin{center}
\includegraphics[width=9cm]{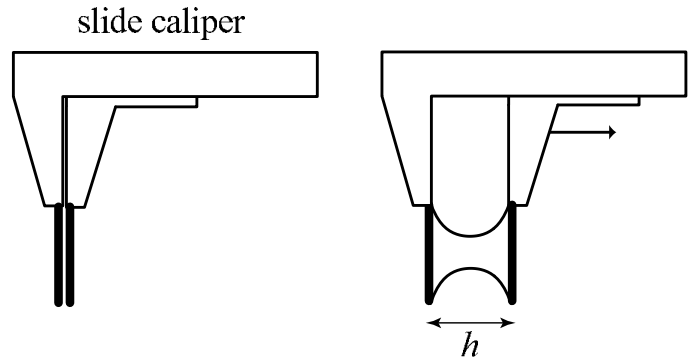}
\caption{\label{fig-5}
Schematic of catenoid formation process.
Soap water is introduced between the rings when $h=0$ (left panel).
The soap-film catenoid forms by pulling a string attached to the movable
part of the caliper (right panel).
The catenoid shape is recorded by a video camera mounted over the slide
caliper.
}
\end{center}
\end{figure}

\begin{figure}
\begin{center}
\includegraphics[width=12cm]{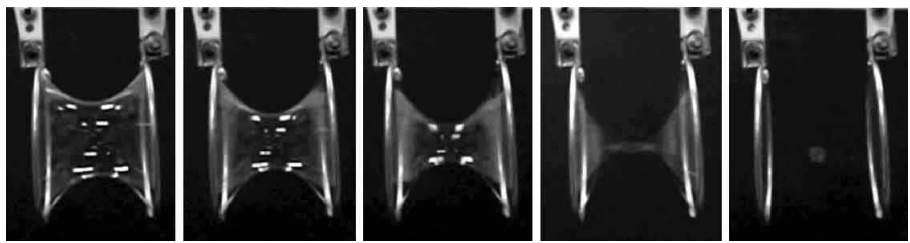}
\caption{\label{fig-6}
Sequential photographs showing the disappearance of the catenoid.
The distance between two rings increases from left to right.
}
\end{center}
\end{figure}

The experimental procedure is as follows.
As indicated in caption of Fig. \ref{fig-5}, when $h=0$, soapy water is
introduced between the rings.
Next, by pulling a string attached to the movable part
of the slide caliper, we can observe the formation of the soap-film
catenoid.
While pulling the string, the detailed behaviour of catenoid is recorded
by a video camera with motion capture (1/120 sec per frame) mounted
above the caliper (Fig. \ref{fig-6}).
Using this apparatus, we observe the catenoid in-situ.

\begin{figure}
\begin{center}
\includegraphics[width=12cm]{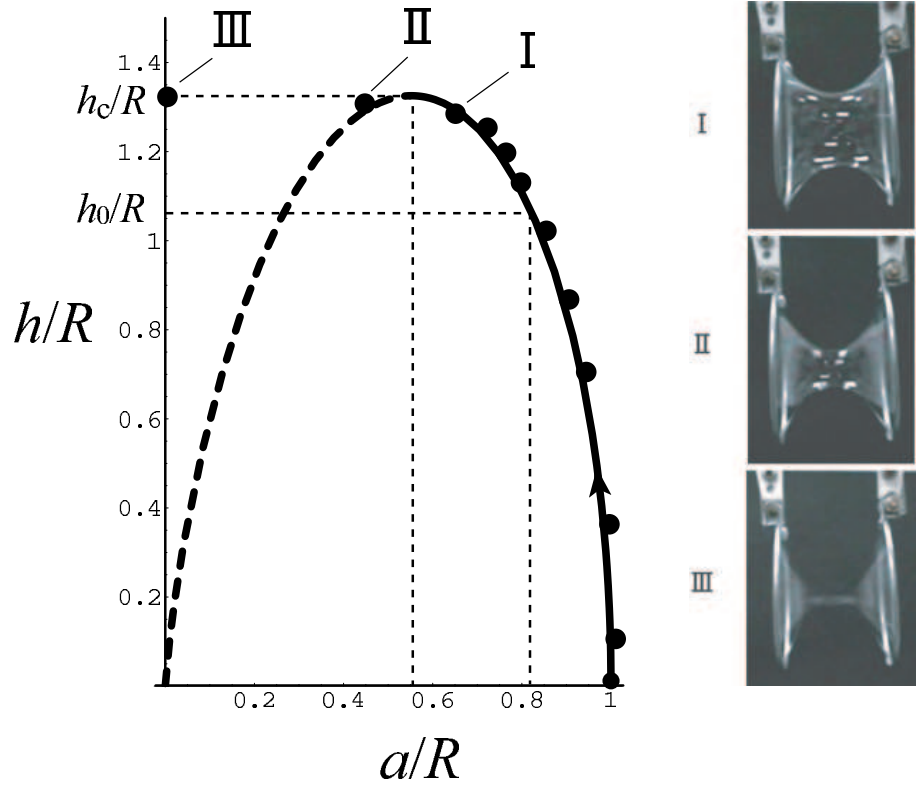}
\caption{\label{fig-7}
The dashed curve and the solid curve in the left panel correspond to the
unstable catenoid and the stable catenoid, respectively.
The dots indicates data acquired from the video images.
In the right panel, photographs corresponding to points I,II,III from
the left panel are shown.
}
\end{center}
\end{figure}

While increasing the distance $h$, we recorded sequential images
from the formation to the rupture of the catenoid, which are
shown in Fig. \ref{fig-6}.
By analysing these photographs, we can measure both the
distance $h$ and neck radius $a$ to generate the graph shown in 
Fig. \ref{fig-7}.

By analyzing photographs, the graph as shown in Fig. \ref{fig-7} can be
depicted.
Starting from $h=0\;(a=R)$, the shape of the soap film varies along the
theoretical curve of a stable catenoid, as shown in Fig. \ref{fig-2}.
As mentioned in Sec. II, the graph of Fig. \ref{fig-3} 
predicts that the stable catenoid transforms into two disks at $h=h_{0}$.
However, in the experiment, the jump from the stable
catenoid to two disks does not occur.
Furthermore, as the distance $h$ approaches the
critical distance $h_{c}=1.33 R$, the catenoid exhibits strange behaviour.
Just before the critical point, we observe for an instant the
unstable catenoid.
The photograph of the unstable catenoid is shown in panel II of 
Fig. \ref{fig-7}.
Immediately, the unstable catenoid collapses, following which
the soap film transforms into two disks.

\begin{figure}
\begin{center}
\includegraphics[width=13cm]{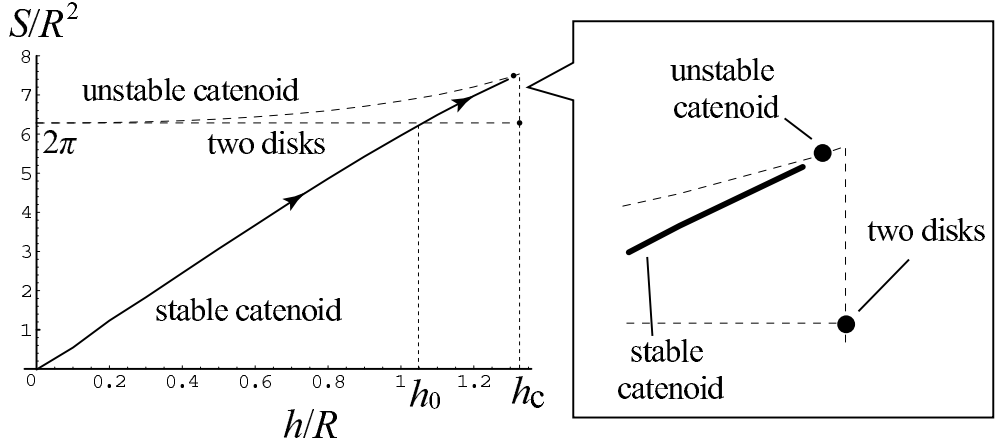}
\caption{\label{fig-8}
Trace of the actual transition of soap film (bold line with arrow in
 left panel and dotted lines in insect).
In the neighbourhood of the critical point, the stable catenoid transforms
for a moment into the unstable catenoid, following which
it vanishes and the soap film forms two disks.
The insect shows a magnified view of this transformation process.
}
\end{center}
\end{figure}

A plot of the actual transition of the soap film is shown
in Fig. \ref{fig-8}.
For $0<h<h_{0}$, soap film occupies the stable catenoid, which is
the absolute minimum for the surface area.
When $h=h_{0}$, the surface area of the stable catenoid is equal to that
of the two disks.
For $h_{0}<h$, the surface area of the stable catenoid is greater than
that of the two disks, which implies that the stable catenoid
corresponds to a local minimum.
For $h_{0}\lesssim h\lesssim h_{\rm c}$, 
the stable catenoid corresponding to a local minimum exists.
Finally, in the neighbourhood of $h\sim h_{c}$, the stable catenoid
transforms into the unstable catenoid, then
immediately collapses and transforms into two disks.

To summarize, the actual sequence followed by the shape of the soap-film
is,
\begin{align}
&{\rm stable\; catenoid\; (absolute\; minimum)}\;\stackrel{\rm (i)}{\to}\;
{\rm stable\; catenoid\; (local\; minimum)}\nonumber\\
&\stackrel{\rm (ii)}{\to}\;
{\rm unstable\; catenoid\; (local\; maximum)}\;\stackrel{\rm (iii)}{\to}\;
{\rm rupture}\;\stackrel{\rm (iv)}{\to}
\;{\rm two\; disks\; (absolute\; minimum)}.\nonumber\\
\end{align}
Using a high quality video camera, we have observed
the interesting behaviours of the soap film.
In the next section, we interpret the behaviour not by
kinematic analysis but by dynamic analysis.

\section{Dynamic interpretation of catenoid}

\begin{figure}
\begin{center}
\includegraphics[width=12cm]{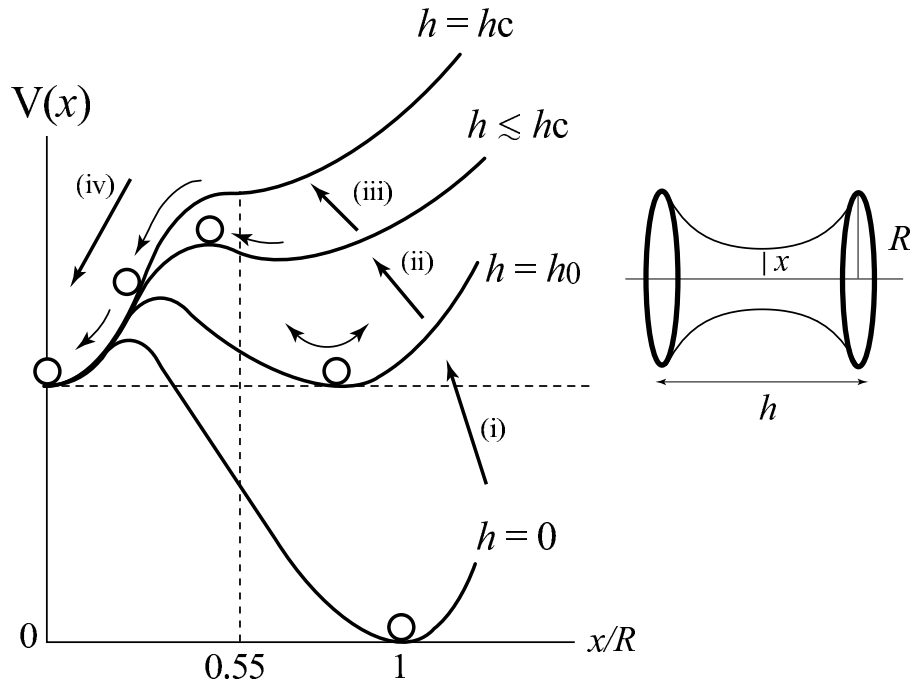}
\caption{\label{fig-9}
The surface energy $V(x)$ of soap film for various values of the
parameter $h$.
Here $x$ represents the minimum catenoid radius.
The circles indicate the positions occupied by the film.
Equation $(5)$ corresponds to
the transitions (i)$\to$(ii)$\to$(iii)$\to$(iv).
}
\end{center}
\end{figure}

To explain the steps (i)-(iv) of (5),
we must consider the surface energy $V(x)$ of a soap film
suspended two rings.
As shown in Fig. \ref{fig-9}, 
the coordinate $x$ is measured mid-way between the two rings when $h$ is
fixed.
By taking into account (5), we can imagine the form of $V(x)$.

Figure \ref{fig-8} indicates that the surface potential energy
$V(x)$ has three extremal points.
If the soap film occupies a stable catenoid or the two disks, then the surface
energy has a local minimal value \cite{paper1}.
The soap film of the unstable catenoid corresponds to a surface
potential energy with the local maximum value.
The illustration of surface energy $V(x)$ is depicted in 
Fig. \ref{fig-9}.
Steps (i)-(iv) of (5) correspond to the transitions shown in
Fig. \ref{fig-9} between surface energy curves.

Because the dynamic behaviour depends on the form of the surface
potential energy, we can see the form of $V(x)$ by observing the soap
film.
The circles in Fig. \ref{fig-9} correspond to the position $x$ of the
soap film (i.e. $x$ corresponds to the neck radius $a$).
The transitions (i)-(iv) of Fig. \ref{fig-9} are interpreted below.

(i) When $h=0$, $V(x)=0$.
At this point, the soap film cannot form a membrane.
Because the neck radius $a=R$, the circle in Fig. \ref{fig-9} is located
at $x=R$, which corresponds to an absolute minimum of $V(x)$.
Upon increasing $h$, the catenoid forms and the
absolute minimum point is raised because the formation of the soap film
increases the surface energy $V$.
The soap film shape is slightly perturbed because the string is pulled
by hand to increase $h$, and this perturbation gives rise to
fluctuations around the extremal point.
However, despite of these fluctuations, the stable catenoid
is in a dynamically stable state because it occupies an absolute minimum
point on the surface energy curve.
As $h$ approaches $h_{0}$, the surface area of stable catenoid
approaches that of the surface area of the two disks.
When $h=h_{0}$, the surface energy of the stable catenoid and
the two disks is the same, and both are absolute minimum values.

(ii) Increasing $h$ increases the absolute minimum point
corresponding to the stable catenoid.
For $h_{0}<h<h_{c}$,
the corresponding circle is in a local minimum point.
In spite of the perturbations, the jump from stable catenoid to
two rings does not occur
because the potential barrier separating the local minimum from the
absolute minimum at $x=0$ prevents the soap
film from transferring to the absolute minimum point which corresponds to
two disks.
This analysis reveals that kinematic analysis of Sec. II
is not correct.

(iii) As $h$ approaches $h_{c}$, the local minimum point in $V(x)$
attains the height necessary to overcome the potential barrier.
Because of the perturbation, at $h\lesssim h_{c}$
the circle can cross over the local maximum point corresponding to the
unstable catenoid, which
we succeeded in photographing (Fig. {\ref{fig-7}}).
In Fig. \ref{fig-9}, this situation corresponds to the circle rolling 
over the top of a hill.
Using a high quality video camera, it is possible to record the catenoid
going over this local maximum in the surface energy.

(iv) The catenoid collapses after crossing over the maximum point, and 
the soap film transfers to the two rings corresponding to the
absolute minimum point a $x=0$.
Referring to Fig. \ref{fig-9}, the interpretation is that
the circle rolls down a steep potential-energy hill.

Thus, the in-situ observations of a catenoid formed by soap film are completely
explained by this dynamic interpretation.

At this point, two comments are in order.
First, the concentration of soapsuds is not accounted for in
the analysis.
Secondly, the speed at which the rings are separated is not taken
into consideration.
In this experiment, a precise measurement of separation speed was
difficult.
Because we took care to keep the separation speed slower than competing
processes (e.g., catenoid collapse), we consider that the speed does not
change the essentials of the interpretation of
the phenomenon.

\section{Summary}

We demonstrate in this article that
the in-situ observation of soap film is suitable as an educational physics
experiment.
In the experiment proposed here, students can observe various
interesting phenomena connected with local extrema. 
The film behaviour is interpreted using dynamic mechanics. 
Through the experiment and the analysis, students have the opportunity
to study many fields, including
differential geometry, minimal surfaces, and dynamical mechanics.
The experimental apparatus is inexpensive and only common materials are
required.
We thus recommend the in-situ observation of soap-film catenoids as an
educational physics experiment.

\section*{Acknowledgements}
The authors gratefully acknowledge 
the financial support of Aichi University of Education.
%
\section*{Appendix: Catenoid equation}

\begin{figure}
\begin{center}
\includegraphics[width=12cm]{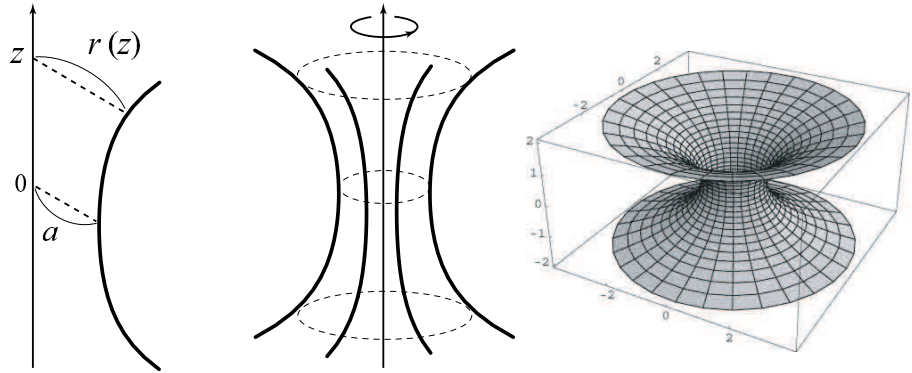}
\caption{\label{fig-1}
A catenary curve with the radius $r(z)$ and the minimum radius $a$
labelled (left panel).
A catenoid is made by rotating the catenary curve around the $z$ axis
(centre panel).
A three-dimensional schematic of a catenoid (right panel, created by using
Mathematica 5.1).
}
\end{center}
\end{figure}

We derive the catenoid equation of Eq. (2).
The cylindrical coordinate in $z$-axial symmetry is represented by 
$\r=(r(z)\cos\phi,r(z)\sin\phi,z)$ in Fig. \ref{fig-1}, where $\phi$ is the
angle around $z$-axis.
The surface area of revolution around $z$-axis is given by
\begin{align}
S=\int \left|\frac{\p\r}{\p z}\times\frac{\p\r}{\p\phi}\right|d\phi dz=
2\pi\int r\sqrt{1+r^{\prime 2}}\;dz\,,
\end{align}
where $\dis\prime=\frac{d}{dz}$.
We can read off the Lagrangian 
$\dis L=r\sqrt{1+r^{\prime 2}}$ from Eq. (6),
then the Euler-Lagrange equation leads to
\begin{align}
\frac{d}{dz}\left(\frac{rr^{\prime}}{\sqrt{1+r^{\prime 2}}}\right)=
\sqrt{1+r^{\prime 2}}\,.
\end{align}
Furthermore, Eq. (7) can be simplified as follows
\begin{align}
\frac{1}{r^{\prime}}
\frac{d}{dz}\left(\frac{r}{\sqrt{1+r^{\prime 2}}}\right)=0\,.
\end{align}
By integrating it, we get
\begin{align}
\frac{r}{\sqrt{1+r^{\prime 2}}}=a\,,
\end{align}
where $a$ is constant.
Thus, we can obtain catenoid equation
\begin{align}
r(z)=a\cosh\frac{z-C}{a}\,,
\end{align}
where $C$ is constant.
The case of $C=0$ is Eq. (2).



\newpage

\end{document}